\documentclass[floatfix,notitlepage,aps,prd,twocolumn,superscriptaddress,showpacs,altaffilletter,preprintnumbers]{revtex4-1}
\usepackage{graphicx}
\usepackage{amsmath}
\usepackage{amssymb}
\usepackage{longtable}
\usepackage[mathscr]{euscript}
\usepackage{lineno}

\usepackage{physics}

\begin{document}

\title{Constraints on Neutrino Lifetime from the Sudbury Neutrino Observatory}

%
\newcommand{\dec}{Deceased}
\newcommand{\alta}{Department of Physics, University of 
Alberta, Edmonton, Alberta, T6G 2R3, Canada}
\newcommand{\chicago}{Department of Physics, University of 
Chicago, Chicago IL} 
\newcommand{\ubc}{Department of Physics and Astronomy, University of 
British Columbia, Vancouver, BC V6T 1Z1, Canada}
\newcommand{\bnl}{Chemistry Department, Brookhaven National 
Laboratory,  Upton, NY 11973-5000}
\newcommand{\carleton}{Ottawa-Carleton Institute for Physics, Department of Physics, Carleton University, Ottawa, Ontario K1S 5B6, Canada}
\newcommand{\carletona}{Department of Physics, Carleton University, Ottawa, Ontario, Canada}
\newcommand{\uog}{Physics Department, University of Guelph,  
Guelph, Ontario N1G 2W1, Canada}
\newcommand{\lu}{Department of Physics and Astronomy, Laurentian 
University, Sudbury, Ontario P3E 2C6, Canada}
\newcommand{\lbnl}{Institute for Nuclear and Particle Astrophysics and 
Nuclear Science Division, Lawrence Berkeley National Laboratory, Berkeley, CA 94720-8153}
\newcommand{\lbla}{ Lawrence Berkeley National Laboratory, Berkeley, CA}
\newcommand{\lanl}{Los Alamos National Laboratory, Los Alamos, NM 87545}
\newcommand{\llnl}{Lawrence Livermore National Laboratory, Livermore, CA}
\newcommand{\lanla}{Los Alamos National Laboratory, Los Alamos, NM 87545}
\newcommand{\oxford}{Department of Physics, University of Oxford, 
Denys Wilkinson Building, Keble Road, Oxford OX1 3RH, UK}
\newcommand{\penn}{Department of Physics and Astronomy, University of 
Pennsylvania, Philadelphia, PA 19104-6396}
\newcommand{\pennx}{Department of Physics and Astronomy, University of 
Pennsylvania, Philadelphia, PA}
\newcommand{\queens}{Department of Physics, Queen's University, 
Kingston, Ontario K7L 3N6, Canada}
\newcommand{\uw}{Center for Experimental Nuclear Physics and Astrophysics, 
and Department of Physics, University of Washington, Seattle, WA 98195}
\newcommand{\uwx}{Center for Experimental Nuclear Physics and Astrophysics, 
and Department of Physics, University of Washington, Seattle, WA}
\newcommand{\uta}{Department of Physics, University of Texas at Austin, Austin, TX 78712-0264}
\newcommand{\triumf}{TRIUMF, 4004 Wesbrook Mall, Vancouver, BC V6T 2A3, Canada}
\newcommand{\ralimp}{Rutherford Appleton Laboratory, Chilton, Didcot, UK} 
\newcommand{\iusb}{Department of Physics and Astronomy, Indiana University, South Bend, IN}
\newcommand{\fnal}{Fermilab, Batavia, IL}
\newcommand{\uo}{Department of Physics and Astronomy, University of Oregon, Eugene, OR}
\newcommand{\hu}{Department of Physics, Hiroshima University, Hiroshima, Japan}
\newcommand{\slac}{Stanford Linear Accelerator Center, Menlo Park, CA}
\newcommand{\mac}{Department of Physics, McMaster University, Hamilton, ON}
\newcommand{\doe}{US Department of Energy, Germantown, MD}
\newcommand{\lund}{Department of Physics, Lund University, Lund, Sweden}
\newcommand{\mpi}{Max-Planck-Institut for Nuclear Physics, Heidelberg, Germany}
\newcommand{\uom}{Ren\'{e} J.A. L\'{e}vesque Laboratory, Universit\'{e} de Montr\'{e}al, Montreal, PQ}
\newcommand{\cwru}{Department of Physics, Case Western Reserve University, Cleveland, OH}
\newcommand{\pnnl}{Pacific Northwest National Laboratory, Richland, WA}
\newcommand{\uc}{Department of Physics, University of Chicago, Chicago, IL}
\newcommand{\mitt}{Laboratory for Nuclear Science, Massachusetts Institute of Technology, Cambridge, MA 02139}
\newcommand{\ucsd}{Department of Physics, University of California at San Diego, La Jolla, CA }
\newcommand{	\lsu	}{Department of Physics and Astronomy, Louisiana State University, Baton Rouge, LA 70803}
\newcommand{\imp}{Imperial College, London, UK}
\newcommand{\uci}{Department of Physics, University of California, Irvine, CA 92717}
\newcommand{\ucia}{Department of Physics, University of California, Irvine, CA}
\newcommand{\suss}{Department of Physics and Astronomy, University of Sussex, Brighton  BN1 9QH, UK}
\newcommand{\sussx}{Department of Physics and Astronomy, University of Sussex, Brighton, UK}
\newcommand{\lifep}{Laborat\'{o}rio de Instrumenta\c{c}\~{a}o e F\'{\i}sica Experimental de
Part\'{\i}culas, Av. Elias Garcia 14, 1$^{\circ}$, 1000-149 Lisboa, Portugal}
\newcommand{\lipx}{Laborat\'{o}rio de Instrumenta\c{c}\~{a}o e F\'{\i}sica Experimental de
Part\'{\i}culas,  Lisboa, Portugal}
\newcommand{\hku}{Department of Physics, The University of Hong Kong, Hong Kong.}
\newcommand{\aecl}{Atomic Energy of Canada, Limited, Chalk River Laboratories, Chalk River, ON K0J 1J0, Canada}
\newcommand{\nrc}{National Research Council of Canada, Ottawa, ON K1A 0R6, Canada}
\newcommand{\princeton}{Department of Physics, Princeton University, Princeton, NJ 08544}
\newcommand{\birkbeck}{Birkbeck College, University of London, Malet Road, London WC1E 7HX, UK}
\newcommand{\snoi}{SNOLAB, Lively, ON P3Y 1N2, Canada}
\newcommand{\snoix}{SNOLAB, Lively,  ON, Canada}
\newcommand{\uba}{University of Buenos Aires, Argentina}
\newcommand{\hvd}{Department of Physics, Harvard University, Cambridge, MA}
\newcommand{\pny}{Goldman Sachs, 85 Broad Street, New York, NY}
\newcommand{\pnv}{Remote Sensing Lab, PO Box 98521, Las Vegas, NV 89193}
\newcommand{\nts}{Nevada National Security Site, Las Vegas, NV}
\newcommand{\psis}{Paul Schiffer Institute, Villigen, Switzerland}
\newcommand{\liverpool}{Department of Physics, University of Liverpool, Liverpool, UK}
\newcommand{\uto}{Department of Physics, University of Toronto, Toronto, ON, Canada}
\newcommand{\uwisc}{Department of Physics, University of Wisconsin, Madison, WI}
\newcommand{\psu}{Department of Physics, Pennsylvania State University,
     University Park, PA}
\newcommand{\anl}{Deparment of Mathematics and Computer Science, Argonne
     National Laboratory, Lemont, IL}
\newcommand{\cornell}{Department of Physics, Cornell University, Ithaca, NY}
\newcommand{\tufts}{Department of Physics and Astronomy, Tufts University, Medford, MA}
\newcommand{\ucd}{Department of Physics, University of California, Davis, CA}
\newcommand{\unc}{Department of Physics, University of North Carolina, Chapel Hill, NC}
\newcommand{\dresden}{Institut f\"{u}r Kern- und Teilchenphysik, Technische Universit\"{a}t Dresden,  Dresden, Germany} 
\newcommand{\isu}{Department of Physics, Idaho State University, Pocatello, ID}
\newcommand{\qmul}{School of Physics and Astronomy, Queen Mary University of London, UK}
\newcommand{\ucsb}{Dept. of Physics, University of California, Santa Barbara, CA}
\newcommand{\cern}{CERN, Geneva, Switzerland}
\newcommand{\utah}{Dept. of Physics, University of Utah, Salt Lake City, UT}
\newcommand{\casa}{Center for Astrophysics and Space Astronomy, University
of Colorado, Boulder, CO}
\newcommand{\susel}{Sanford Underground Research Laboratory, Lead, SD}  
\newcommand{\ntu}{Center of Cosmology and Particle Astrophysics, National Taiwan University, Taiwan}
\newcommand{\berlin}{Institute for Space Sciences, Freie Universit\"{a}t Berlin,
Leibniz-Institute of Freshwater Ecology and Inland Fisheries, Germany}
\newcommand{\bhsu}{Black Hills State University, Spearfish, SD} 
\newcommand{\queensa}{Dept.\,of Physics, Queen's University, 
Kingston, Ontario, Canada} 
\newcommand{\aasu}{Dept.\,of Chemistry and Physics, Armstrong  State University, Savannah, GA}
\newcommand{\ucb}{Physics Department, University of California at Berkeley, Berkeley, CA 94720-7300}
\newcommand{\ucbx}{Physics Department, University of California at Berkeley, and Lawrence Berkeley National Laboratory, Berkeley, CA}
\newcommand{\mcgill}{Physics Department, McGill University, Montreal, QC, Canada}
\newcommand{\columbia}{Columbia University, New York, NY}
\newcommand{\rhul}{Dept. of Physics, Royal Holloway University of London, Egham, Surrey, UK}
\newcommand{\ubama}{Department of Physics and Astronomy, University of Alabama, Tuscaloosa, AL}
\newcommand{\kit}{Institut f\"{u}r Experimentelle Kernphysik, Karlsruher Institut f\"{u}r Technologie, Karlsruhe, Germany}
\newcommand{\winnipeg}{Department of Physics, University of Winnipeg, Winnipeg, Manitoba, Canada}
\newcommand{\kwantlen}{Kwantlen Polytechnic University, Surrey, BC, Canada}
\newcommand{\cea}{CEA-Saclay, DSM/IRFU/SPP, Gif-sur-Yvette, France}
\newcommand{\sunysb}{Laufer Center, Stony Brook University, Stony Brook, NY}
\newcommand{\rock}{Rock Creek Group, Washington, DC}
\newcommand{\rcnp}{Research Center for Nuclear Physics, Osaka, Japan}
\newcommand{\usd}{University of South Dakota, Vermillion, SD}
\newcommand{\lancaster}{Physics Department, Lancaster University, Lancaster, UK}
\newcommand{\potsdam}{GFZ German Research Centre for Geosciences, Potsdam, Germany}
\newcommand{\kirchhoff}{Ruprecht-Karls-Universit\"{a}t Heidelberg, Im Neuenheimer Feld 227, Heidelberg, Germany}
\newcommand{\continuum}{Continuum Analytics,  Austin, TX}
\newcommand{\gsu}{Dept. of Physics, Georgia Southern University, Statesboro, GA}
\newcommand{\pelmorex}{Pelmorex Corp., Oakville, ON} 
\newcommand{\usaid}{Global Development Lab, U.S. Agency for International Development, Washington DC}
    

\affiliation{\alta}
\affiliation{\ucb}
\affiliation{\ubc}
\affiliation{\bnl}
\affiliation{\carleton}
\affiliation{\uog}
\affiliation{\lu}
\affiliation{\lbnl}
\affiliation{\lifep}
\affiliation{\lanl}
\affiliation{\lsu}
\affiliation{\mitt}
\affiliation{\oxford}
\affiliation{\penn}
\affiliation{\queens}
\affiliation{\snoi}
\affiliation{\uta}
\affiliation{\triumf}
\affiliation{\uw}
\author{B.~Aharmim}\affiliation{\lu}
\author{S.\,N.~Ahmed}\affiliation{\queens}
\author{A.\,E.~Anthony}\altaffiliation{Present address: \usaid}\affiliation{\uta}
\author{N.~Barros}\altaffiliation{Present address: \pennx}\affiliation{\lifep}
\author{E.\,W.~Beier}\affiliation{\penn}
\author{A.~Bellerive}\affiliation{\carleton}
\author{B.~Beltran}\affiliation{\alta}
\author{M.~Bergevin}\altaffiliation{Present address: \llnl}\affiliation{\lbnl}\affiliation{\uog}
\author{S.\,D.~Biller}\affiliation{\oxford}
\author{R.~Bonventre}\affiliation{\ucb}\affiliation{\lbnl}
% ...
			
\collaboration{SNO Collaboration}
\noaffiliation

\date{\today}

\begin{abstract}
The long baseline between the Earth and the Sun makes solar neutrinos an excellent test beam for exploring possible neutrino decay.
The signature of such decay would be an energy-dependent distortion of the traditional survival probability which can be fit for using well-developed and high precision analysis methods.
Here a model including neutrino decay is fit to all three phases of $^8$B solar neutrino data taken by the Sudbury Neutrino Observatory.
This fit constrains the lifetime of neutrino mass state $\nu_2$ to be ${>8.08\times10^{-5}}$ s/eV at $90\%$ confidence.
An analysis combining this SNO result with those from other solar neutrino experiments results in a combined limit for the lifetime of mass state $\nu_2$ of ${>1.04\times10^{-3}}$ s/eV at $99\%$ confidence.
\end{abstract}

\maketitle

\section{\label{Intro} Introduction}
Nuclear reactions in the core of the Sun produce electron flavor neutrinos at rates which can be predicted by solar models. 
Neutrinos produced in the solar $^8$B reaction propagate to the Earth and are detected as electron flavor neutrinos with a probability, $P_{ee}$, of roughly $1/3$, with the remainder converted to $\nu_{\mu}$ or $\nu_{\tau}$.
Analysis of data from the Sudbury Neutrino Observatory (SNO) \cite{sno} and Super-Kamiokande (SK) \cite{superk} has shown the origin of this $1/3$ survival probability to be due to mixing of neutrino states with finite mass that are distinct from the flavor states in which neutrinos are produced and interact.
Many other experiments \cite{homestake,sage,gallex,kamland,borexino} have made precision measurements of solar neutrino fluxes probing the rich physics of neutrino mixing, and are consistent with this conclusion.
With the discovery of finite neutrino mass comes the possibility that neutrinos may be unstable and could decay to some lighter particle.

Neutrino decay was first explored as a possible explanation for the less-than-unity survival probability of electron flavor neutrinos \cite{bachall_stable} (the so-called ``solar neutrino problem''). 
The flavor-tagging \cite{SNOCC} and flavor-neutral \cite{d2o} detection channels of the SNO detector unambiguously demonstrated by a flavor-independent measurement of the neutrino flux that the solar neutrino problem was not due to neutrino decay.
Even though neutrino decay is now known not to be the dominant effect behind the solar neutrino problem,  solar neutrinos make an excellent test beam for investigating neutrino decay as a second-order effect.

Astrophysical and cosmological observations provide strong constraints on radiative decay of neutrinos (see references in the Particle Data Group (PDG) review \cite{pdg}), however constraints on nonradiative decay are much weaker.
The energy ranges of solar neutrinos and the baseline between the Sun and the Earth make solar neutrinos a strong candidate for setting constraints on nonradiative decays \cite{beacombell}. 
As such, we consider only nonradiative decays where any final states would not be detected as active neutrinos \cite{beacombell,berryman,picoreti,choubey,joshipura,bandyopadhyay2003,raghavan}.
The signal of nonradiative neutrino decay is an energy-dependent disappearance of neutrino flux that can be extracted with a statistical fit to solar neutrino data.

Previous analyses of neutrino lifetime~\cite{berryman,picoreti} utilizing published SNO fits~\cite{3phase} were limited because the polynomial survival probabilities from~\cite{3phase} do not well capture the shape distortion of neutrino decay.
Additionally, the previously published fits assumed the total flux was conserved, i.e. they inferred $P_{ea}$, the probability of detecting a solar neutrino as a $\nu_\mu$ or $\nu_\tau$ neutrino at Earth, from the constraint $P_{ee} + P_{ea} = 1$ which does not apply in a decaying scenario.
Both points are addressed in this analysis by implementing and fitting to a model including neutrino decay that independently calculates $P_{ee}$ and $P_{ea}$.

Precise measurements of neutrino mixing parameters from KamLAND~\cite{kamland} and Daya Bay~\cite{dayabay} along with improved theoretical predictions for the $^8$B flux~\cite{serenelli} reduce the uncertainty in the underlying solar neutrino model.
Using these constraints, a dedicated fit is performed to SNO data where the electron neutrino survival probabilities, $P_{ee}$ and $P_{ea}$, are calculated directly as an energy-dependent modification to the standard Mikheyev - Smirnov - Wolfenstein (MSW)~\cite{wolfenstein,mikheyev} survival probability.

This paper is organized as follows. In Section \ref{detector} we discuss the SNO detector.  
Section \ref{theory} reviews the theoretical basis of the measurement.
Section \ref{analysis} presents the analysis technique, a likelihood fit of the solar neutrino signal that includes a neutrino decay component.
The results are presented in Section \ref{results}, and Section \ref{conclusion} concludes.

\section{\label{detector} The SNO Detector}
\begin{figure}
\includegraphics[width=\columnwidth]{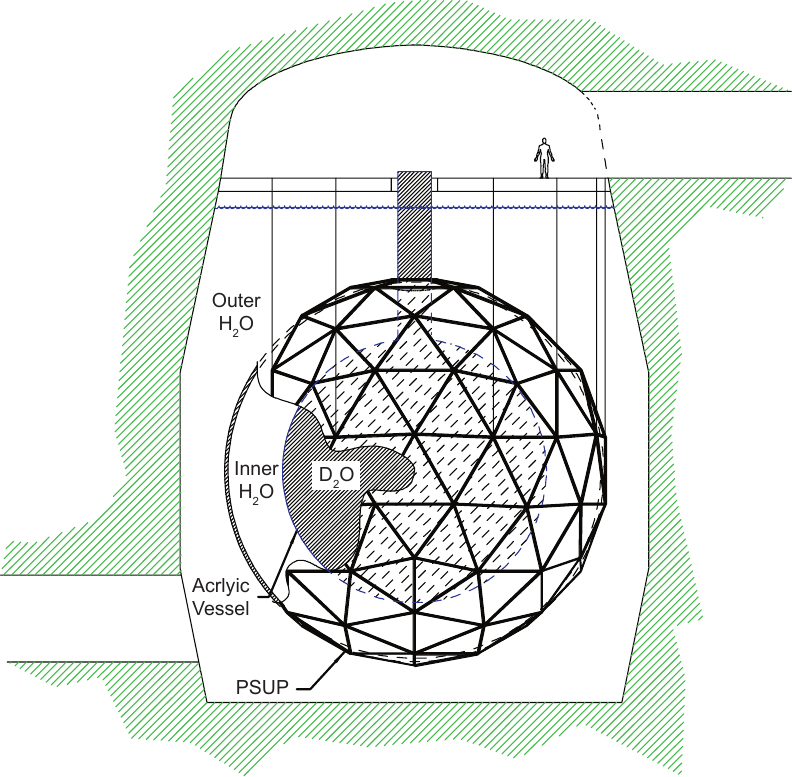}
\caption{\label{fig:det} The SNO detector~\cite{3phase}.}
\end{figure}

The Sudbury Neutrino Observatory was a heavy water Cherenkov detector located at a depth of 2100 m (5890 m.w.e.) in Vale's Creighton mine near Sudbury, Ontario.
The detector utilized an active volume of 1000 metric tons of heavy water (D$_2$O) contained within a 12 m diameter spherical acrylic vessel (AV).
The AV was suspended in a volume of ultrapure light water (H$_2$O) which acted as shielding from radioactive backgrounds.
This ultrapure water buffer contained 9456 8-inch photomultiplier tubes (PMTs) attached to a 17.8 m diameter geodesic structure (PSUP).
These PMTs recorded Cherenkov light produced by energetic particles in the active volume.
The effective coverage of the PMTs was increased to 55\% \cite{Jelley} by placing each PMT inside a non-imaging reflective light concentrator.  
A schematic diagram of the detector is shown in Figure \ref{fig:det}.

SNO was sensitive to three neutrino interaction channels:
\begin{center}
\begin{tabular}{llclrll}
$\nu$ & + d & $\rightarrow$ & p + n + $\nu$ &-- 2.22 MeV & (NC) & , \\
$\nu_e$ & + d & $\rightarrow$ & p + p + e$^-$ &-- 1.44 MeV & (CC) & , \\
$\nu$ & + e$^-$ & $\rightarrow$ & $\nu$ + e$^-$ & & (ES) & .
\end{tabular} 
\end{center}
The neutral current interaction (NC) couples to neutrinos of all flavors equally and allowed an unambiguous measurement of the total active neutrino flux.
The charged current (CC) and elastic scattering (ES) interactions couple exclusively (CC) or preferentially (ES) to the electron flavor neutrino, which allowed the solar electron neutrino survival probability to be measured.

SNO operated in three phases, which differed in sensitivity to neutrons, and hence to the NC interaction.
Phase I was the baseline detector described above in which neutrons were detected via the 6.25 MeV $\gamma$-ray released after capturing on deuterons.
Phase II increased the neutron capture efficiency using the higher capture cross section of $^{35}$Cl by adding NaCl to the D$_2$O.
In addition to the increased cross section, the neutron capture on $^{35}$Cl resulted in a cascade of $\gamma$-rays summing to a higher energy of 8.6 MeV, better separating this signal from radioactive backgrounds.
Phase III added a Neutral Current Detector (NCD) array inside the active volume for an independent measure of neutron production inside the detector. 
These NCDs were high purity nickel tubes containing $^3$He gas, and they were instrumented to utilize the $^3$He as a proportional counter for thermal neutrons~\cite{ncd}.
For Phase III only there are two sources of detector data: the PMT array data as in Phase I and Phase II and the NCD array data. 
As these datasets are treated differently in analyses, the PMT data from Phase III will be referred to simply as Phase III with the NCD data being Phase IIIb.
A combined analysis of Phase I and II data led to the first low energy measurement of the electron neutrino survival probability \cite{leta}. 
That analysis was later extended to incorporate Phase III data \cite{3phase}, and the analysis described in this paper was based on the analysis described in \cite{3phase}.

SNO developed a highly detailed microphysical simulation of the detector called SNOMAN \cite{detectorpaper}.  
This software could be configured to exactly reflect the experimental conditions at any particular time (for example, the values of the trigger settings during a particular run), allowing accurate Monte Carlo reproduction of the data.
Monte Carlo simulations of the various signal and background events generated with statistics equivalent to many years of livetime were used extensively in this analysis.
For a detailed description of this simulation package, see \cite{3phase}.

\section{\label{theory} Neutrino Decay for $^8$B Solar Neutrinos}
This section provides the theoretical background for the analysis.  
We begin by reviewing ordinary solar neutrino oscillation and the MSW effect before introducing the effects of possible neutrino decay.

\subsection{Neutrino oscillation}

Neutrinos are produced and interact in the flavor basis, $\ket{\nu_{\alpha}}$ where $\alpha = {e,\mu,\tau}$, however these are not eigenstates of the vacuum Hamiltonian, whose eigenstates (the eigenstates with definite mass) we denote as $\ket{\nu_i}$ where $i = {1,2,3}$. 
The flavor basis is related to the mass basis by the Pontecorvo-Maki-Nakagawa-Sakata (PMNS) matrix $U_{\alpha i}$ as follows:
\begin{equation}
\ket{\nu_{\alpha}} = \sum_i U^*_{\alpha i} \ket{\nu_i}.
\end{equation}
The free evolution of these states is most easily represented in the mass basis, as these are eigenstates of the Hamiltonian:
\begin{equation}
\bra{\nu_i}H_{0}\ket{\nu_j} = \frac{1}{2E}\begin{bmatrix}
m_1^2 & 0 & 0 \\
0 & m_2^2 & 0 \\
0 & 0 & m_3^2
\end{bmatrix}.
\end{equation}
The survival probability for flavor state $\ket{\nu_\alpha}$ to be detected as flavor state $\ket{\nu_\beta}$ at some later time after free evolution for a distance L is therefore
\begin{equation}
P_{\alpha\beta} = \left|\braket{\nu_\beta(t)}{\nu_\alpha}\right|^2 = \left|\sum_i U^*_{\alpha i} U^{}_{\beta i} e^{- i m_i^2 L/(2E)} \right|^2.
\end{equation}

\subsection{The MSW effect}

The MSW~\cite{wolfenstein,mikheyev} effect proposes that the coherent forward scattering of electron flavor neutrinos off of electrons in a material adds a potential energy, $V_e$, to electron flavor neutrinos which depends on the local electron density, $n_e$:
\begin{equation}
V_e = \sqrt{2} G_F n_e.
\end{equation}
Written in the mass basis the Hamiltonian including this effect, $H_{MSW}$, then takes the form
\begin{equation}
H_{MSW} = H_{0} + U\begin{bmatrix}
V_e & 0 & 0 \\
0 & 0 & 0 \\
0 & 0 & 0
\end{bmatrix}U^\dagger.
\end{equation}
Notably the evolution of the states in the presence of matter is now much more complicated since the eigenstates now depend on the electron density.

It is useful to introduce the matter mass basis, $\ket{\nu_{mi}(V_e)}$, consisting of eigenstates of the Hamiltonian $H_{MSW}$ at a particular electron potential $V_e$.
Note that if the electron density is zero, $H_{MSW}$ reduces to $H_0$. Therefore $\ket{\nu_{mi}(0)} \rightarrow \ket{\nu_{i}}$.
In many cases the variation of $V_e$ is slow enough that the evolution is adiabatic, meaning some state $\ket{\nu(t)}$ has a constant probability to be one of the instantaneous matter mass states at a later time. 
\begin{equation}
\left| \braket{\nu_{mi}(V_e(0))}{\nu(0)} \right|^2 = \left| \braket{\nu_{mi}(V_e(t))}{\nu(t)} \right |^2
\end{equation}

An adiabatic approximation is made in this analysis as with previous SNO analyses~\cite{3phase}. 
Knowing where in the Sun a neutrino is produced (or more precisely the electron density at the production point), one can calculate the eigenstate composition for as long as the adiabatic condition is satisfied.
Once the neutrino reaches the solar radius, vacuum propagation dominates. 
As vacuum propagation does not change the mass state composition of a state, the neutrinos that arrive at Earth have the same mass state composition as those exiting the Sun.
Due to the large distance between the Earth and the Sun, these mass state fluxes can be assumed to be incoherent once they arrive at Earth, and any regeneration of coherence in the Earth is ignored.
Therefore, the arrival probability $\phi_i$ of neutrino mass state $\nu_i$ at Earth due to electron neutrinos produced at an electron potential $V_e$ in the Sun in the presence of the MSW effect can be calculated as
\begin{equation}
\phi_i = \left| \braket{\nu_{m i}(V_e)}{\nu_e} \right|^2.
\label{rawflux}
\end{equation}
The analytic expression for this value is non-trivial and in practice $H_{MSW}$ is numerically diagonalized in the flavor basis to find $\bra{\nu_{m i}(V_e)}$ at a particular $V_e$ value and compute this projection.

\subsection{Modeling a neutrino decay signal}
The flux of a particular mass state, $i$, could have some lifetime associated with it, $\tau_i$, representing the decay of neutrinos of that mass state.
Since the actual neutrino masses are currently unknown, the lifetime may be represented by an effective parameter, $k_i$, scaled by the mass of the state:
\begin{equation}
k_i = \frac{\tau_i}{m_i}.
\end{equation}
Since the Earth-Sun distance is quite large compared to the solar radius, any decay within the sun will be ignored, and decay is only considered while propagating in vacuum from the Sun to the Earth.
Here, we consider nonradiative decay to some non-active channel \cite{beacombell}, which manifests as disappearance of a mass state.
Therefore, the arrival probability, $\psi_i$, of a neutrino mass state at Earth in the presence of neutrino decay can be given as 
\begin{equation}
\psi_i \approx  e^{-L / (E k_i)} \phi_i =  e^{-L / (E k_i)} \left| \braket{\nu_{m i}(V_e)}{\nu_e} \right|^2
\label{decay}
\end{equation}
where $L$ is the radius of the Earth's orbit (1 AU) and $E$ is the energy of the neutrino.
Survival probabilities for electron and muon/tau flavor neutrinos may then be recovered using the PMNS matrix in the usual way:
\begin{equation}
\begin{array}{rcl}
P_{ee} & = & \sum_i \psi_i |U_{ie}|^2  \\
P_{ea} & = & \sum_i \psi_i |U_{i\mu}|^2 + \psi_i |U_{i\tau}|^2.
\end{array}
\label{survive}
\end{equation}

\subsection{Decay of $^8$B solar neutrinos}

\begin{figure}
\includegraphics[width=0.9\columnwidth]{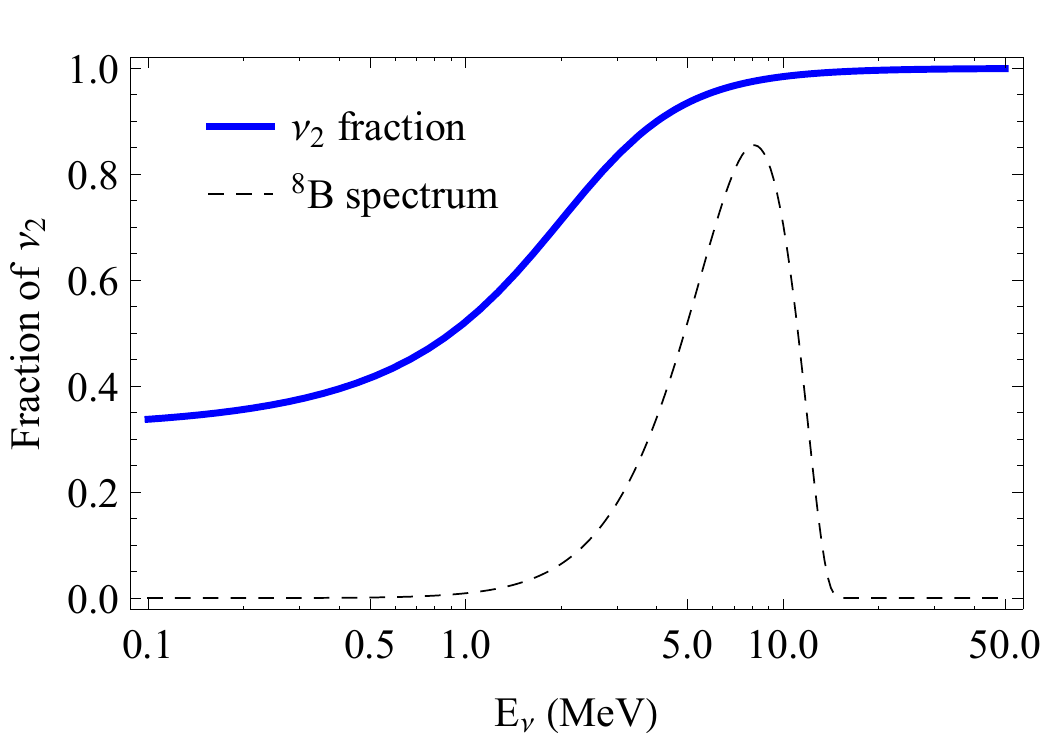}
\caption{\label{nu2flux}The fraction of solar neutrino flux that, due to the MSW effect, is mass state $\nu_2$ is shown here in a solid line. The cross section weighted $^8$B energy spectrum is shown with a dashed line to guide the eye. The $\nu_2$ state dominates over the energy range where $^8$B neutrinos can be detected.}
\end{figure}

Figure \ref{nu2flux} shows the fraction of mass state $\nu_2$ in the total neutrino flux as a function of energy.
Considering the cross section weighted $^8$B neutrino energy spectrum, one finds that less than $4\%$ of the detected flux is not mass state $\nu_2$.
As such, SNO data is dominated by $\nu_2$ neutrinos.
This analysis is insensitive to decay of mass states $\nu_1$ or $\nu_3$, and the lifetimes $k_1$ and $k_3$ are assumed to be infinite.

\begin{figure}
\includegraphics[width=0.9\columnwidth]{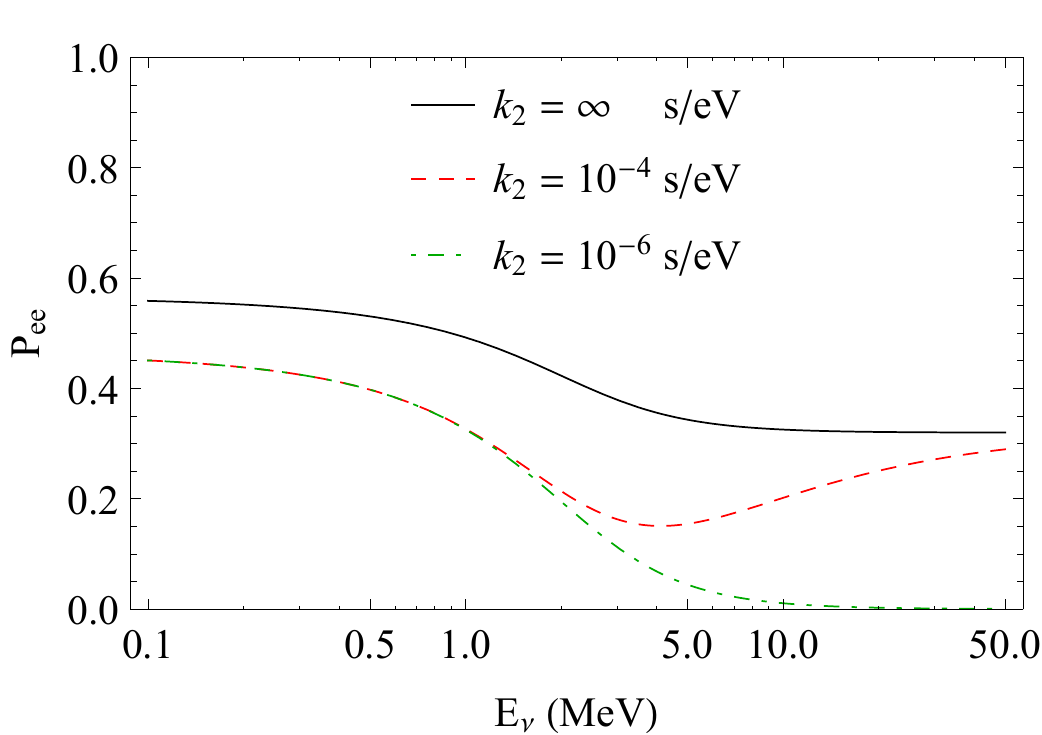}
\caption{\label{decayexamples}Shown here in dashed lines are survival probabilities of electron neutrinos, $P_{ee}$, for various values of mass state $\nu_2$ lifetime ($k_2$) demonstrating the energy-dependent distortion being fit for. Existing limits are near $k_2 = 10^{-4}$ s/eV. The solid line shows the survival probability with no neutrino decay.}
\end{figure}

The signal to be fit is therefore an energy-dependent flux disappearance due to the decay of mass state $\nu_2$ neutrinos. 
This energy dependence is distinct from the MSW effect, allowing an energy-dependent likelihood fit to distinguish between them.
In the formalism presented here, decay of mass state $\nu_2$ is entirely described by the lifetime parameter $k_2$.
Examples of $P_{ee}$ for various values of $k_2$ are shown Figure \ref{decayexamples}.

\section{\label{analysis} Analysis}
We performed a likelihood fit over all three phases of SNO data for a finite neutrino lifetime, $k_2$, as defined in the previous section. 
This analysis built on the 3-phase SNO analysis \cite{3phase} and the methods are briefly summarized here for completeness but can be found in detail in the previous publication. 
For each fit, many parameters were floated with constraints. 
These parameters include background rates, neutrino mixing parameters, and the nominal $^8$B flux. 
Systematic uncertainties found not to be strongly correlated with the solar neutrino signal were handled with a shift and refit procedure. 
For the final result, a likelihood profile for the parameter $k_2$ was generated and used to set a lower bound for that parameter.  See the following sections for more detail.

\subsection{Data selection}
Data selection proceeds in a number of steps.  
The data are organized in time periods called runs, and the first step is to select runs with nominal detector conditions.  
This analysis uses the same run list developed for the full analysis of all three phases of the SNO data \cite{3phase}.

There is also an event-level selection within each run.
These cuts remove instrumental backgrounds, muons, and muon followers from the dataset.  
Again, for this analysis we use the same reconstruction corrections, data cleaning, and high-level cuts used in \cite{3phase} for identifying physics events.

We define a region of interest for the analysis in terms of effective recoil electron kinetic energy $T_{\textrm{eff}}$ and radial position $r$, requiring $r <$ 5.5 m, and $3.5 \textrm{ MeV} < T_{\textrm{eff}} < 20 \textrm{ MeV} $.  
Phase III data is included with a higher range of energies, $6.0 \textrm{ MeV} < T_{\textrm{eff}} < 20 \textrm{ MeV}$, as in previous SNO analyses~\cite{3phase}, since low energy backgrounds were not as well understood in that phase.

\subsection{Blindness}
The data from all three phases of SNO were reblinded during the development of the analysis.
The fit itself was developed on a statistical ensemble of Monte Carlo datasets.
Once the analysis was finalized, the data were unblinded in two stages.
The fit was first run on a one-third statistical subsample, to verify that it behaved as expected on real data, before proceeding to fit the full dataset.

\subsection{Fit}
We developed a binned likelihood fit that combines all three phases of SNO data.
For Phases I and II, we perform a fit in four observable quantities: energy, volume-weighted radius ($\rho = r^3/r_{AV}^3$), solar angle, and isotropy ($\beta_{14}$).  
For Phase III data, we perform a fit in three observable quantities: energy, radius, and solar angle.
To incorproate Phase IIIb data, we use a constraint from the earlier pulse shape analysis \cite{3phase} that determined the number of NCD events that could be attributed to neutrino interactions.
For each of these components, the binning of the observable quantities used was that in \cite{3phase}.

For each class of signal and background events in a phase, a probability distribution function (PDF) with the correct dimensions for that phase is produced using Monte Carlo events. The likelihood of the data being described by a weighted sum of the PDFs for each class of signal and background is maximized by minimizing the negative logarithm of this likelihood with MINUIT \cite{minuit}.
The construction of this likelihood function is identical to what is described in the SNO 3-phase analysis \cite{3phase} with one exception: the polynomial survival probability from previous SNO analyses is replaced with the survival probability parameterized by the physical quantities described in Section \ref{theory}.

\subsection{Solar Signal}

The following sections discuss the inputs to modeling the flux of $^8$B solar neutrinos as detected by SNO.


\subsubsection{Standard Solar Model}
The neutrino model implemented here uses the radial distribution of electron density and radial distribution of the $^8$B neutrino flux calculated in the BS05(OP) Standard Solar Model (SSM)~\cite{bs05op}.
Uncertainties in these values are not quoted in the original source and are therefore not considered in this fit.
These predictions are expected to be uncorrelated with $\nu_2$ decay as they are not determined with neutrino measurements.

As earth-bound measurements of the solar neutrino flux would be biased by neutrino decay, a theoretical prediction for the $^8$B flux is required.
Serenelli's most recent prediction~\cite{serenelli} yields a $^8$B flux of $5.88\times10^{6}$ cm$^{-2}$s$^{-1}$ with $11\%$ uncertainty which is used as a prior in this fit.
For reference the flux from BS05(OP)~\cite{bs05op} is $5.69\times10^6$ cm$^{-2}$s$^{-1}$.

\subsubsection{Neutrino Mixing\label{mixing}}
Neutrino mixing parameters taken from KamLAND~\cite{kamland} and Daya Bay~\cite{dayabay} are reproduced in Table \ref{tbl:mixing_params}. 
Parameters from KamLAND and Daya Bay were used in this analysis to avoid biasing the result by using values correlated with previous SNO analyses.
As these measurements were done with neutrinos produced on Earth, they are expected to be uncorrelated with effects of $\nu_2$ decay given existing constraints on neutrino decay.
The current limit on $k_2$ constrains it to be $> 7.2\times10^{-4}$ s/eV~\cite{picoreti} which means at length scales comparable to the diameter of the Earth, the maximum flux fraction lost by $\nu_2$ decay for a $10$ MeV neutrino is given by
$1-e^{-2R_{earth}/(E k_2)} \approx 6\times10^{-6}$.
Such a small fractional loss would have negligible impact on values quoted for mixing parameters.
These parameters and their central values are used as priors and floated during the fit.

\begin{table}
\centering
\begin{tabular}{c|c|c}
Parameter & Value & Ref \\ \hline
$ \Delta m ^2 _{21} $ & $7.58^{+0.14}_{-0.13}(stat)^{+0.15}_{-0.15}(syst) \times 10^{-5}$ eV$^2$ & \cite{kamland} \\ \hline
$ \tan^2 \theta_{12} $ & $0.56^{+0.10}_{-0.07}(stat)^{+0.10}_{-0.06}(syst) $ & \cite{kamland} \\ \hline
$ |\Delta m ^2 _{32}| $ & $2.45\pm0.06(stat)\pm0.06(syst) \times 10^{-3}$ eV$^2$ & \cite{dayabay} \\ \hline
$ \sin^2 2\theta_{13} $ & $0.0841\pm0.0072(stat)\pm0.0019(syst)$ & \cite{dayabay} \\ \hline
$ \sin^2 \theta_{23} $ & $0.5^{+0.058}_{-0.062}$ & \cite{superkth23} \\ 
\end{tabular}
\caption{
\label{tbl:mixing_params}
Reproduced here are the mixing parameters used in this analysis taken from Daya Bay~\cite{dayabay}, Super-Kamiokande~\cite{superkth23}, and KamLAND~\cite{kamland} results.
}
\end{table}

\subsection{Backgrounds\label{sec:backgrounds}}
Besides instrumental backgrounds, which can be easily removed with cuts based on event topology, the main sources of background events are radioactive backgrounds and atmospheric neutrino interactions. 
A summary of the sources of these and other backgrounds is given in this section.

For Phases I and II, radioactive decays of $^{214}$Bi (from uranium and radon chains) and $^{208}$Tl (from thorium chains) produce both $\beta$-particles and $\gamma$-rays with high enough energies to pass event selection criteria.
In Phase III the lower energy bound was high enough to exclude these backgrounds.
The inner D$_2$O, acrylic vessel, and outer H$_2$O volumes are treated as separate sources of radioactive decays due to differing levels of contamination.
The PMT array is another source of radioactivity and, despite its increased distance from the fiducial volume, is the dominant source of low-energy backgrounds.

Relevant to all three phases, $\gamma$-rays above $2.2$ MeV may photodisintegrate deuterium resulting in a neutron background.
Radon daughters present on the acrylic vessel since construction result in additional neutron backgrounds from ($\alpha$,$n$) reactions on carbon and oxygen in the acrylic.
In Phase II the addition of NaCl resulted in a $^{24}$Na background from neutron captures on $^{23}$Na.
$^{24}$Na decay produces a $\gamma$-ray with high enough energy to photodisintegrate a deuteron, increasing the neutron background in Phase II.
In Phase III the addition of the NCD array inside the acrylic vessel brought additional radioactive backgrounds. 
Primarily this resulted in an increase of photodisintegration events throughout the detector. Two NCDs with higher levels of radioactivity were treated separately in the analysis.

Additional backgrounds include solar $hep$ neutrinos and atmospheric neutrinos. 
The $hep$ neutrinos have a higher endpoint than $^8$B neutrinos, however the predicted flux is approximately a thousand times less \cite{bs05op}. 
The flux of $hep$ neutrinos is fixed to the standard solar model rate in this analysis.
Atmospheric neutrinos also have a relatively low flux, and the rate is fixed to results from previous SNO analyses \cite{leta}.

Finally, there is a class of instrumental background that tends to reconstruct on the acrylic vessel. 
For Phase III these instrumentals are easily cut in event selection as they were well separated from physics events in the $\beta_{14}$ parameter. Near the lower energy threshold in Phase I and Phase II these events were not as well separated in $\beta_{14}$ resulting in some contamination \cite{leta}, and this event class was therefore included in the fit for Phase I and Phase II. 

For further details on how backgrounds were included in the fit and which in-situ and ex-situ constraints were used, see Appendix B in \cite{3phase}.

\subsection{\label{syst} Systematics}

Parameters that shift, rescale, or affect the resolution of observables used in the fit are treated as systematic uncertainties. 
Other systematic uncertainties include: parameters that control the shape of the analytic PDF for PMT $\beta-\gamma$ backgrounds, photodisintegration efficiency, and neutron capture efficiencies.

The neutron capture efficiency was found to be strongly correlated with the neutrino parameters and is floated in the fit to correctly account for correlations with the final results.

Less correlated parameters that are well constrained by the data, such as the parameters for the analytic PDF for PMT $\beta-\gamma$ backgrounds, are scanned as an initial step.
Each of these systematic parameters is scanned independently with other systematic parameters held fixed while profiling out all floated parameters. 
This scan produces a likelihood profile, which is fit by an asymmetric Gaussian to determine the central value and uncertainty of the scanned parameter.
After a parameter is scanned, its central value and uncertainty is updated to the fit result before scanning the next parameter.
This process is repeated until the central values for each parameter stabilize to ensure the global minimum is found.
The final central values are retained and fixed during MINUIT minimization, and the impact of their uncertainty on the uncertainty of floated parameters is evaluated with the shift-and-refit procedure described below.

The least correlated parameters that are not well constrained by the data are fixed to predetermined nominal values during the fit, and the impact of their uncertainty on each floated parameter is evaluated with a shift-and-refit method.

The shift-and-refit method draws $10^6$ sets of systematic parameters from their respective asymmetric Gaussian distributions.
The fit is then re-run many times with the systematic parameters fixed to each of the generated sets.
This produces distributions of fitted values for the parameters floated in the fit.
The widths of these distributions are taken to represent the systematic uncertainty on the floated parameters.

For a full listing of the systematic uncertainties and how they were handled, see Appendix B in \cite{3phase}.

\subsection{Bias and pull testing}
Significant testing was done on Monte Carlo datasets to ensure the statistical robustness of the fit.
For all tests in this section, the solar signal was generated with an assumed $k_2$ value of $10^{-4}$ s/eV to test the sensitivity near existing limits, and all other parameters were chosen by randomly sampling the prior distributions for each fake dataset. 

The fit was run on each dataset including only the signals and backgrounds for that stage.
The fitted values of all floated parameters were recorded and used to produce bias and pull distributions. 
No significant bias was found, and pull widths were found to be consistent with expectations.

\section{\label{results} Results}

\begin{figure}
\includegraphics[width=\columnwidth]{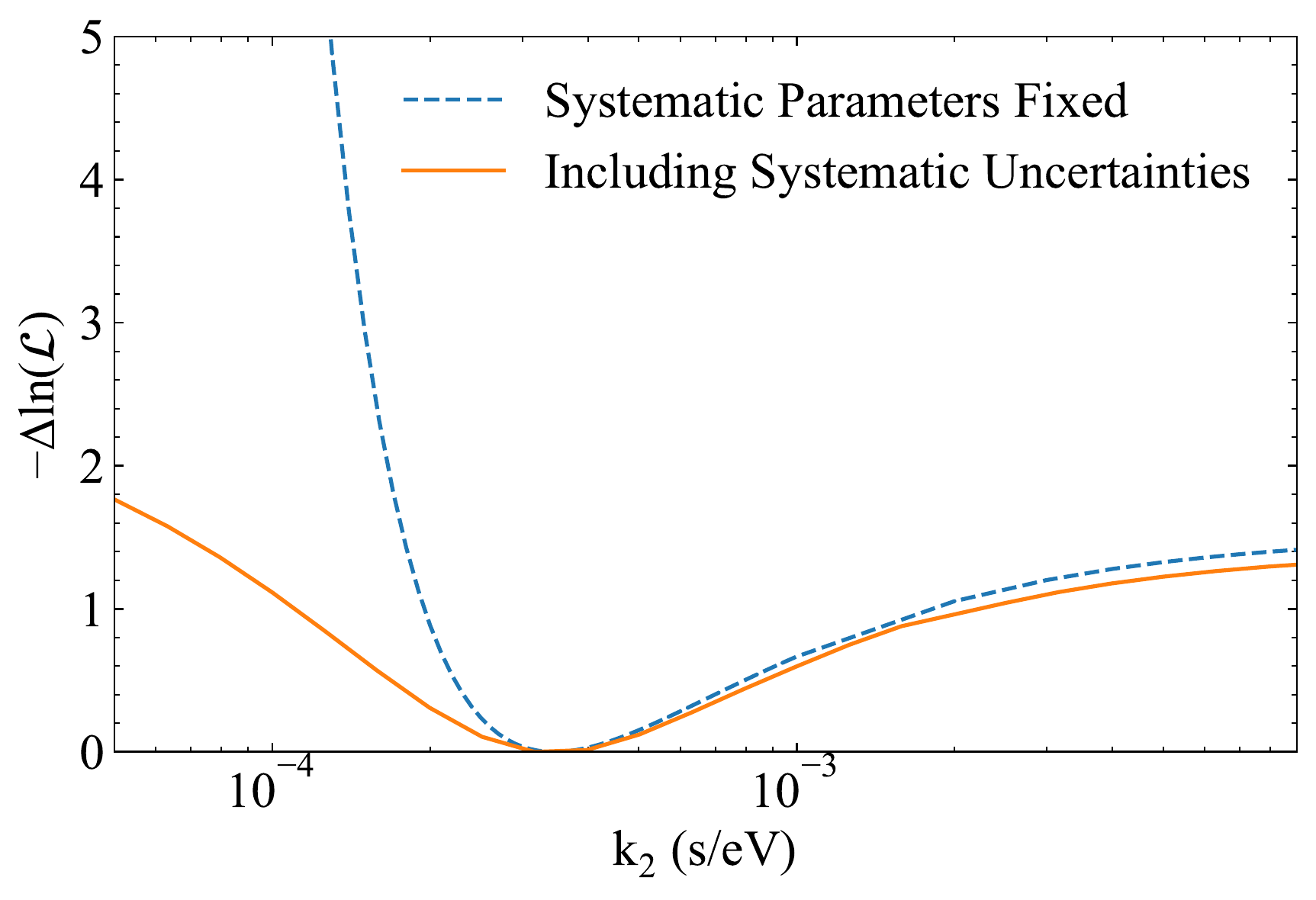}
\caption{\label{k2_scan} Shown here is the likelihood scan of the mass state $\nu_2$ lifetime $k_2$. This is shown both with systematic parameters fixed, and after incorporating systematic uncertainties using a shift-and-refit method.}
\end{figure}

Figure \ref{k2_scan} shows the likelihood profile of $k_2$ both with the systematic parameters fixed to central values and with the systematic uncertainties included. 
The likelihood profile incorporating systematic uncertainties is generated by assuming the shape of the likelihood profile does not change as the systematic parameters vary, but rather simply shifts according to the shift in the fitted value of $k_2$ from the shift-and-refit method.
Therefore, the systematic uncertainties are included by shifting the fixed systematic profile by each shift in the shift-and-refit distribution for $k_2$ and averaging the likelihood at each point. 

A shallow minimum at $3.45^{+5.50}_{-1.68}\times10^{-4}$ s/eV is found, however the upper uncertainty is consistent with infinite lifetime at confidences greater than $85\%$, meaning this analysis is not a significant measurement of neutrino decay. 
Using Wilks' theorem~\cite{wilks1938}, a lower bound for $k_2$ can be set at $k_2>8.08\times10^{-5}$ s/eV at $90\%$ confidence.

\subsection{Comparison to previous SNO analyses}
The best fit $^8$B neutrino flux from this analysis is $6.08^{+0.47}_{-0.47}$(stat.)$^{+0.21}_{-0.22}$(syst.)$\times10^6$ cm$^{-2}$s$^{-1}$ and has slight tension with results of the previous SNO 3-phase analysis: $5.23^{+0.16}_{-0.16}\times10^6$ cm$^{-2}$s$^{-1}$ \cite{3phase} where statistical and systematic uncertainties have been combined.
With $k_2$ fixed to infinite lifetime this analysis results in a $^8$B neutrino flux of $5.22^{+0.16}_{-0.16}\times10^6$ cm$^{-2}$s$^{-1}$, again with systematic and statistical uncertainties combined, that is in very good agreement with previous results.
The uncertainty with $k_2$ allowed to float is much larger due to the additional freedom of neutrino decay in the model and the fact that the lifetime is strongly anti-correlated with the flux.
These two parameters are not degenerate only because the effect of neutrino decay is energy-dependent, and this fit to the neutrino energy spectrum can capture that effect.
To that end we expect the uncertainty on the $^8$B neutrino flux from this analysis to be larger than previous analyses.

\subsection{Combined analysis results \label{combined_analysis}}

Any experiment measuring a solar flux can be compared to a standard solar model to constrain neutrino lifetimes.
Likelihood profiles of $k_2$ generated for other solar experiments can be combined with the profile from this analysis to arrive at a global limit.
Particularly, experiments sensitive to lower energy solar neutrinos, such as the $pp$ or $^7$Be solar neutrinos, can provide strong constraints on $k_2$ as the $L/E$ for these neutrinos is greater than for the $^8$B neutrinos.

To incorporate the results from other experiments, the measured flux reported by an experiment assuming a flux of only electron neutrinos (i.e. $P_{ee} = 1$), $\Phi_e$, is converted to an total inferred flux, $\Phi_T$, by way of a neutrino model that predicts $\overline{P_{ee}}$ and $\overline{P_{ea}}$, the average survival probabilities for that flux, and the relative cross sections, $\sigma_a / \sigma_e$, where $\sigma_e$ is the cross section for electron-flavor neutrinos and $\sigma_a$ the cross section for all other neutrino flavors (i.e. $\nu_\mu$ and $\nu_\tau$)
\begin{equation}
\Phi_T = \left(\overline{P_{ee}} + \overline{P_{ea}} \frac{\sigma_a}{\sigma_e}\right)^{-1}\Phi_e.
\label{flux_conv}
\end{equation}
The neutrino decay model described in Section \ref{theory} is used, and the averaging is done over the flux-appropriate standard solar model production regions (electron density) \cite{bs05op} and experiment-appropriate cross section weighted neutrino spectra \cite{bs05op}. 

This $\Phi_T$ can be directly compared to standard solar model predictions with the following likelihood term
\begin{equation}
-\mathrm{ln}({\mathscr L}) = \frac{(\Phi_T - \Phi_{SSM})^2}{2 (\sigma_T^2+\sigma_{SSM}^2)}
\label{combolike}
\end{equation}
where $\sigma_T$ and $\sigma_{SSM}$ are the uncertainties on the inferred flux, $\Phi_T$, and standard solar model flux, $\Phi_{SSM}$.
The mass state $\nu_1$ lifetime, $k_1$, is a free parameter in the fit and profiled over in producing the final limit on $k_2$, as lower energy solar neutrinos may contain significant fractions of $\nu_1$.
The mass state $\nu_3$ lifetime, $k_3$, remains fixed to infinity, as all solar neutrinos contain negligible amounts of $\nu_3$.
The neutrino mixing parameters are constrained as described in Section \ref{mixing} and allowed to float.

Following this methodology, a profile for $k_2$ is generated using:
Super-K~\cite{superkiv}, KamLAND~\cite{kamland8b} and Borexino~\cite{borexino8b} $^8$B results;
Borexino~\cite{borexino7be} and KamLAND~\cite{kamland7be} $^7$Be results;
the combined gallium interaction rate from GNO, GALLEX, and SAGE~\cite{sagecombo};
and the chlorine interaction rate from Homestake~\cite{homestake}.
For both chlorine and gallium, the predicted interaction rate is computed following the procedure in Section V of \cite{sagecombo}, but using the neutrino model and mixing constraints used elsewhere in this paper. 
These predicted rates were compared to the measured rates with likelihood terms analogous to Equation \ref{combolike}.

\begin{figure}
\includegraphics[width=\columnwidth]{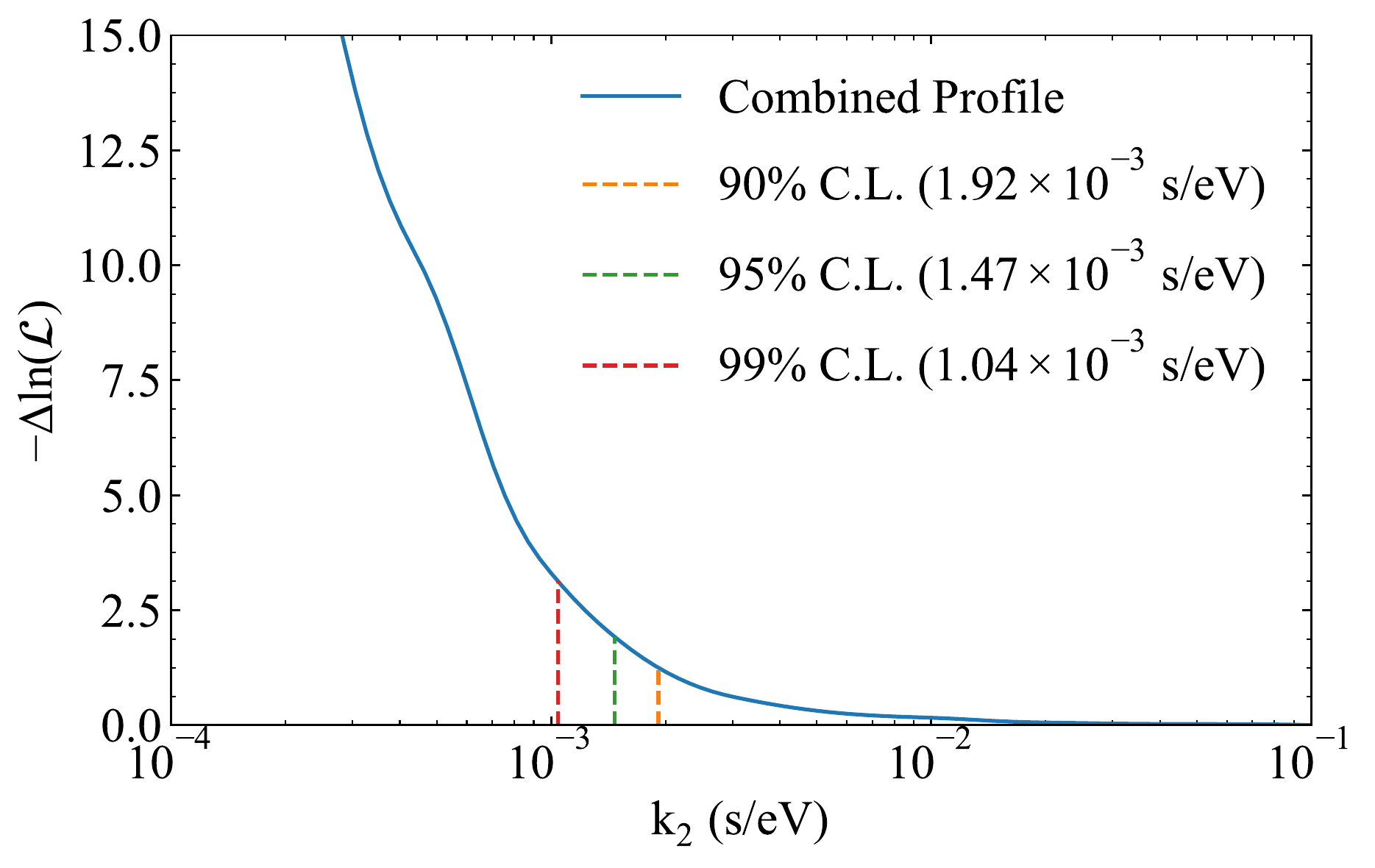}
\caption{\label{combined_scan} Shown here is the combined likelihood profile including the SNO result from this analysis and flux constraints from other solar experiments as described in Section \ref{combined_analysis}.}
\end{figure}

The final profile, combined with this analysis of SNO data, is shown in Figure \ref{combined_scan} and constrains $k_2$ to be ${>1.04\times10^{-3}}$ s/eV at $99\%$ confidence.

\section{\label{conclusion} Conclusion}
Neutrinos are known to have mass, allowing for potential decays to lighter states.
However, analyses of solar neutrino data assuming the MSW solution to the solar neutrino problem are consistent with a non-decaying scenario.
By analyzing the entire SNO dataset, using a model that predicts the survival probability of electron-type solar neutrinos allowing for the decay of mass state $\nu_2$, we were able to set a limit on the lifetime of neutrino mass state $\nu_2$: ${k_2>8.08\times10^{-5}}$ s/eV at $90\%$ confidence.
Combining this with measurements from other solar experiments results in a new best limit of ${k_2>1.04\times10^{-3}}$ s/eV at $99\%$ confidence.
The improvement from the previous limit, $k_2>7.2\times10^{-4}$ s/eV at $99\%$ confidence \cite{picoreti}, was driven by the inclusion of additional solar flux measurements and updated analyses of previously considered experiments. \vspace{16pt}

\begin{acknowledgments}
This research was supported by: Canada: Natural Sciences and Engineering Research Council, Industry Canada, National Research Council, Northern Ontario Heritage Fund, Atomic Energy of Canada, Ltd., Ontario Power Generation, High Performance Computing Virtual Laboratory, Canada Foundation for Innovation, Canada Research Chairs program; US: Department of Energy Office of Nuclear Physics, National Energy Research Scientific Computing Center, Alfred P. Sloan Foundation, National Science Foundation, the Queen’s Breakthrough Fund, Department of Energy National Nuclear Security Administration through the Nuclear Science and Security Consortium; UK: Science and Technology Facilities Council (formerly Particle Physics and Astronomy Research Council); Portugal: Funda\c{c}\~{a}o para a Ci\^{e}ncia e a Tecnologia.  
This research used the Savio computational cluster resource provided by the Berkeley Research Computing program at the University of California, Berkeley (supported by the UC Berkeley Chancellor, Vice Chancellor for Research, and Chief Information Officer).
We thank the SNO technical staff for their strong contributions.  We thank INCO (now Vale, Ltd.) for hosting this project in their Creighton mine.
\end{acknowledgments}

\bibliography{citations}

\end{document}